\RequirePackage{fix-cm}
\documentclass{svjour3}
\pdfoutput=1
\usepackage{graphicx}
\usepackage{amsmath}
\usepackage{amsfonts}
\usepackage{amssymb}
\usepackage{bm}
\usepackage{color}
\usepackage[linesnumbered]{algorithm2e}
\SetKw{Require}{Require}
\SetKw{LogicalAnd}{and}
\SetKw{LogicalOr}{or}
\usepackage{multirow}

\begin{document}

\title{Stable algorithm for event detection in event-driven
  particle dynamics}

\author{Marcus N. Bannerman \and Severin Strobl \and Arno Formella \and
  Thorsten P\"oschel}

\institute{Marcus N. Bannerman \at
  School of Engineering\\
  University of Aberdeen\\
  Fraser Noble Building\\
  Aberdeen, AB24 3UE\\
  United Kingdom
  \and
  Severin Strobl \and Thorsten P\"oschel \at
  Institute for Multiscale Simulation\\
  Friedrich-Alexander-Universit\"at Erlangen-N\"urnberg\\
  Erlangen\\
  Germany
  \and
  Arno Formella \at Department of Computer Science\\
  Universidad de Vigo\\
  Ourense\\
  Spain
}

\date{Received: date / Accepted: date}


\maketitle

\begin{abstract}
  Event-Driven Particle Dynamics is a fast and precise method to
  simulate particulate systems of all scales.  In this work it is
  demonstrated that, despite the high accuracy of the method, the
  finite machine precision leads to simulations entering invalid
  states where the dynamics are undefined. A general event-detection
  algorithm is proposed which handles these situations in a stable and
  efficient manner. This requires a definition of the dynamics of
  invalid states and leads to improved algorithms for event-detection
  in hard-sphere systems.  \keywords{DEM; event-driven; molecular
    dynamics; hard sphere; collision detection;}
\end{abstract}


\section{Introduction}
Event-Driven Particle-Dynamics (EDPD) is the oldest particle
simulation technique~\cite{Alder_Wainwright_1959} and has found
application in a wide range of fields, from predicting vapor-liquid
equilibria~\cite{Cui_Elliott_2002} to the design of granular vibration
dampers~\cite{Bannerman_etal_2011b}. Although typically used to
simulate simple particle models such as the hard-sphere, the EDPD
technique remains a general approach to particle simulations as
potentials can be discretized to accurately approximate more
conventional model systems, such as
Lennard-Jonesium~\cite{Chapela_etal_2010}, or directly fit to physical
data~\cite{Vahid_etal_2008}. Introducing the {\em coefficient of
  restitution}, EDPD algorithms can also be used to simulate systems
of dissipatively interacting particles, such as granular flows.

Hard-sphere EDPD algorithms in particular are often much more
efficient (sometimes by orders of magnitude) than ``soft'' models
which must solve Newton's equation of motion using time-stepping
numerical integration techniques~\cite{Haile_1997}. Disregarding
machine precision, EDPD algorithms solve the dynamics of
discontinuous-potential models (such as hard spheres) analytically and
do not suffer from errors due to the finite time step always used in
numerical quadrature~\cite{Haile_1997}.

Despite these advantages, there remain numerical difficulties in the
implementation of the EDPD algorithm due to the finite precision of
floating point calculations originating from the machine precision.
Small numerical errors in the detection and processing of events can
cause the simulation to enter states where the dynamics is undefined.
Interestingly, this ambiguity in the dynamics of invalid states has
also led to some difficulties in theoretical treatments in the
past~\cite{Dorfman_Ernst_1989}. These difficulties are not discussed
in the earliest EDPD implementations~\cite{Alder_Wainwright_1959} as
they are relatively rare and frequently resolve themselves; however,
for large systems and/or large simulation times, one must provide
rules to handle such situations. In addition, there are systems, such
as dissipative (granular) gases, which are prone to
clustering~\cite{McNamara_Young_1994} such that even for small numbers
of particles these finite-precision errors deteriorate until the
simulation must be halted.  Modified algorithms have been proposed to
combat these difficulties but current solutions are complex and fail
in certain cases~\cite{Poschel_Schwager_2005,Reichardt_Wiechert_2007}
or modify the system dynamics in an undesired
way~\cite{Luding:1998,DeltourBarrat:1997}.

In this work the difficulties of event-detection in EDPD simulations
are outlined and a general algorithm for stable event-detection is
proposed.  In Sec.~\ref{sec:EDPDalgorithm}, the basic event-driven
algorithm is outlined for the hard-sphere model. The origin of invalid
states is then introduced in Sec.~\ref{sec:eventerrors}, before an
improved version of the event-detection algorithm is presented in
Sec.~\ref{sec:improvedalgorithm}. Finally, in
Sec.~\ref{sec:bouncingball}, a more complex example of a bouncing ball
is used to illustrate that the dynamics of invalid states must be
defined, and demonstrates the extension of the stable algorithm for
hard spheres to fixed boundaries.

\section{\label{sec:EDPDalgorithm}Basic event-driven algorithm for
identical hard spheres}

Considering only conservative pairwise interactions of pairs of
particles $i$ and $j$, located at $\vec{r}_i$ and $\vec{r}_j$ respectively, summing
up the total force, $\vec{F}_i$, acting on a particle $i$ yields
\begin{align}
  \vec{F}_i=\sum_{j\neq i}^N \vec{F}_{ij} = -\sum_{j\neq i}^N \nabla
  \phi(\vec{r}_{ij})\,; \quad \vec{r}_{ij}\equiv\vec{r}_{i}-\vec{r}_{j}
\end{align}
where the sum is over all $N$ particles in the system, excluding self
interactions and $\phi$ is the interaction energy. In traditional
time-stepping simulations, the total force on each particle is inserted
into Newton's equation of motion and numerically integrated to determine
all particle positions and velocities at later times.

In contrast, discrete potentials preclude the use of numerical
quadrature to solve Newton's equation of motion. For example, the
fundamental property of hard-sphere particles is that they cannot
deform one another, that is, their interaction energy reads
\begin{equation}
  \label{eq:intpot}
  \phi^\text{HS}\left(\vec{r}_{ij}\right) =
  \begin{cases}
    \infty & \text{if~} \left|\vec{r}_{ij}\right| < \sigma \\
    0 & \text{if~} \left|\vec{r}_{ij}\right| \ge \sigma
  \end{cases}
\end{equation}
where $\sigma$ is their collision diameter. Consequently, the
particles move on ballistic trajectories except when two particles
reach the distance $\sigma$. The divergence of
$\phi^\text{HS}(\sigma)$ implies that at this point, there is an
infinite repelling force which in turn implies that the duration of
the interaction approaches zero, that is, at this point the velocities
alter instantaneously. Performing a momentum and energy balance over
two colliding hard spheres~\cite{Poschel_Schwager_2005} yields the
collision rule for the evolution of the particle velocity,
$\vec{v}_i$:
\begin{equation}
  \label{eq:collchangeElastic}
  \vec{v}_i^{\,\prime} =
  \vec{v}_i-\frac{2m_j}{m_i+m_j}
  \left(\hat{\vec{r}}_{ij}\cdot\vec{v}_{ij}\right)
  \hat{\vec{r}}_{ij}\,;\quad
  \vec{v}_{ij}\equiv\vec{v}_i-\vec{v}_j\,;\quad
  \hat{\vec{r}}_{ij}\equiv\frac{\vec{r}_{ij}}{\left|\vec{r}_{ij}\right|}
\end{equation}
where the primes denote post-collision values.

From here follows the main idea of EDPD: instead of numerically
integrating Newton's equation of motion, EDPD maps the dynamics of the
many-particle system to a sequence of instantaneous pairwise
interactions. The handling of these two-particle interactions relies
on pre-computed collision operators,
e.g. Eq.~\eqref{eq:collchangeElastic}. Thus, the EDPD algorithm for
hard spheres can be outlined as follows:
\begin{enumerate}
\item The simulation is started with certain initial conditions for
  the positions and velocities of the particles. Obviously, due to
  Eq.~\eqref{eq:intpot}, any legal initial state requires that the
  particles must not overlap one another.
\item Each possible pairing of particles is tested to determine if and
  when a collision is encountered, and the results are used to
  construct a list of all possible future events. This list is then
  sorted to determine the earliest event, which is the only event in
  the list which is guaranteed to occur. If a particle has multiple
  events occuring at the same instant, an ambigutity in the event
  order is introduced; however, it is often implicitly assumed that
  the execution order in these cases is unimportant and these effects
  are not discussed here.
\item \label{step:propagate}The particle positions are propagated
  along ballistic trajectories until the time of the earliest event.
\item The velocities of the two colliding particles are updated
  according to the collision rule, Eq.~\eqref{eq:collchangeElastic}.
\item Any events in the future event list which involve either of the
  colliding particles are updated and the list is re-sorted to
  determine the next event to be processed.
\item Check for conditions to terminate the simulation, e.g. the total
  real time or number of events processed.
\item Continue with step~\ref{step:propagate}.
\end{enumerate}
In contrast to ordinary time-stepping MD algorithms, EDPD progresses
irregularly in time. It jumps from one event to the next and, thus, is
{\em event-driven}. The basic algorithm outlined above contains the
basic ingredients of an EDPD algorithm, albeit it would be unstable
due to precision errors and of time complexity ${\cal
  O}\left(N^2\right)$ per event processed.  Indeed, there is a range
of methods used to accelerate these
calculations~\cite{Rapaport_1980,Lubachevsky:1991}, including the use
of neighbor lists~\cite{Alder_Wainwright_1959}, the delayed states
algorithm~\cite{Jefferson_1985}, calendar priority
queues~\cite{Paul_2007}, and other
optimizations~\cite{Marin:1993,ShidaAnzai:1992,ShidaYamada:1995} which
reduce the calculation costs to constant time (${\cal
  O}\left(1\right)$) per event. Although the cost of simulating each
event is now independent of system size it should be noted that, for
equilibrium systems, the number of events to be processed per unit of
simulation time still scales as $O(N)$. Nevertheless, common to all
algorithms is that the primary cost of simulation arises from the
detection of events.

While the algorithmic challenge of EDPD is the bookkeeping of the list
of future events, the numerical challenge is the computation of the
times of the events and it is the latter which is the subject of this
paper. We will show that na\"{\i}ve algorithms will result in invalid
states and eventually failure of the algorithm due to unavoidable
numerical errors resulting from the finite machine precision.

These problems will be explored using the prototypical discrete
potential, the hard sphere. We wish to point out that EDPD of hard
spheres is not restricted to conservative interactions: Dissipative
collisions may be characterized by the coefficient of normal
restitution defined as the ratio of the post-collisional relative
normal velocity of the particles and the corresponding pre-collisional
value,
\begin{equation}
  \label{eq:epsdef}
  \varepsilon\equiv -\frac{\vec{v}_{ij}^{\,\prime} \cdot
  \vec{r}_{ij}}{\vec{v}_{ij}\cdot \vec{r}_{ij}}\,.
\end{equation}
In general, $\varepsilon$ is a function of the impact velocity and
material properties. It may be analytically obtained by solving Newton's
equation of motion for an isolated pair of colliding particles with the
assumption of a certain interaction force, e.g.~\cite{Schwager:2008,Mueller:2011}. Alternatively, $\varepsilon$ may be
obtained experimentally, e.g.~\cite{Montaine:2011}. The corresponding
collision rule is obtained from the conservation of momentum and angular
momentum and from the loss of kinetic energy quantified by the
coefficient of restitution:
\begin{align}\label{eq:collchange}
  \vec{v}_i^{\,\prime} = \vec{v}_i-\frac{m_j}{m_i+m_j}(1 + \varepsilon)
  \left(\hat{\vec{r}}_{ij}\cdot\vec{v}_{ij}\right)\hat{\vec{r}}_{ij}\,.
\end{align}

From the point of view of physics, the main difference between
integrating Newton's equations of motion using a time-stepping
algorithm for a standard model, such as the Hertz potential, and a
EDPD simulation using hard-spheres is the duration of collisions. This
difference has some subtle consequences leading to limitations of the
applicability of the hard sphere models~\cite{Mueller:2012} which may
be partially overcome~\cite{Mueller:2013}.

\section{\label{sec:eventerrors} Event calculation errors}

\begin{figure}
  \begin{center}
    \includegraphics[width=\columnwidth,clip]{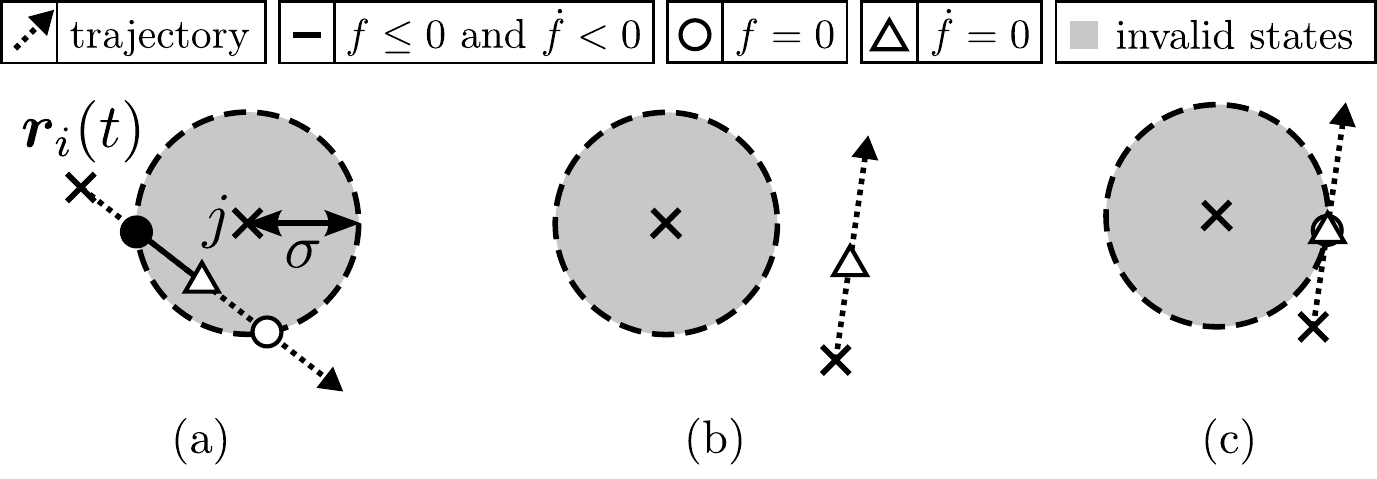}
  \end{center}
  \caption{\label{fig:hard_sphere_roots}
	The three classes of trajectories for a pair of hard-sphere
	particles, (a) colliding, (b) passing, and (c) glancing.  The dotted
	line represents the trajectory of particle $i$ relative to the
	center of particle $j$. The dashed line indicates the border of the
	invalid state/infinite energy region, shaded in gray.  Roots of
	Eq.~\eqref{eq:hs-quadratic}, corresponding to inter-particle
	separations of $\sigma$, are marked with circles. A filled circle
	indicates the impacting root of the overlap function $f$ to be
	introduced in Eq.~\eqref{eq:hs-overlapfunc}, and also marks the
	start of a section of the trajectory which satisfies the stable
	algorithms conditions for an interaction (solid line). Roots of the
	time derivative of the overlap function
	(Eq.~\eqref{eq:hs-overlapfunc-deriv}) are marked with triangles.}
\end{figure}

During the population and updates of the future event list, pair of
particles must be tested to determine if and when they collide. For
identical hard spheres, the equations of motion must be solved to find
if the particle pair approaches to a distance equal to the interaction
diameter, $\sigma$. Assuming that the particles are under identical
acceleration by an external field such as gravity, this detection of
events becomes a search for the time intervals $\Delta t$ which satisfy
\begin{equation}
  \left|\vec{r}_{ij}(t+\Delta t)\right| = \left|\vec{r}_{ij}(t)+\Delta
  t\,\vec{v}_{ij}(t)\right| = \sigma\,.
\end{equation}
Squaring of this expression simplifies it to a quadratic equation in
$\Delta t$
\begin{equation}
  \label{eq:hs-quadratic}
  \Delta t^2\,\vec{v}_{ij}^{\,2} + 2\,\Delta t\,\vec{v}_{ij}\cdot\vec{r}_{ij} +
  \vec{r}_{ij}^{\,2} - \sigma^2 = 0\,,
\end{equation}
where all variables are evaluated at the current time $t$. If this
quadratic equation does not have a real root, the particles do not
come into contact, otherwise there are two roots and the earliest time
root corresponds to the collision.  Fig.~\ref{fig:hard_sphere_roots}
sketches the three classes of trajectories corresponding to zero, two,
and one (degenerated) solution of Eq.~\eqref{eq:hs-quadratic}.

\begin{figure}[htp]
  \begin{center}
    \includegraphics[width=0.7\columnwidth,clip]{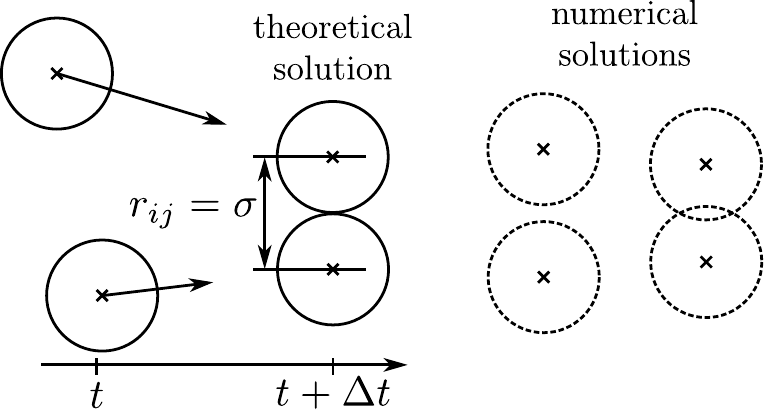}
  \end{center}
  \caption{\label{fig:overlap} An exaggerated illustration of the
	effects of finite precision on the execution of events. Ideally, the
	particles are in contact at the time of the collision; however, due
	to the limited precision the particles are either separated by a
	small gap or end up in a slightly overlapped state.}
\end{figure}

When computing the roots of Eq.~\eqref{eq:hs-quadratic}, small
numerical errors accumulate during the calculation due to the finite
precision of floating point mathematics.  Careful implementations can
minimize these errors through algorithmic improvements or through
arbitrary precision floating-point libraries~\cite{Granlund_2013} but
complete elimination would require exact real
arithmetic~\cite{Mehlhorn_etal_2013} which is relatively
computationally expensive. These errors alter the prediction of the
positions of the colliding particles at the time of impact, which
causes the particles to either numerically overlap or stand a small
distance apart at the time of impact (see Fig.~\ref{fig:overlap}).

\begin{figure}[htp]
  \begin{center}
    \includegraphics[width=0.6\columnwidth,clip]{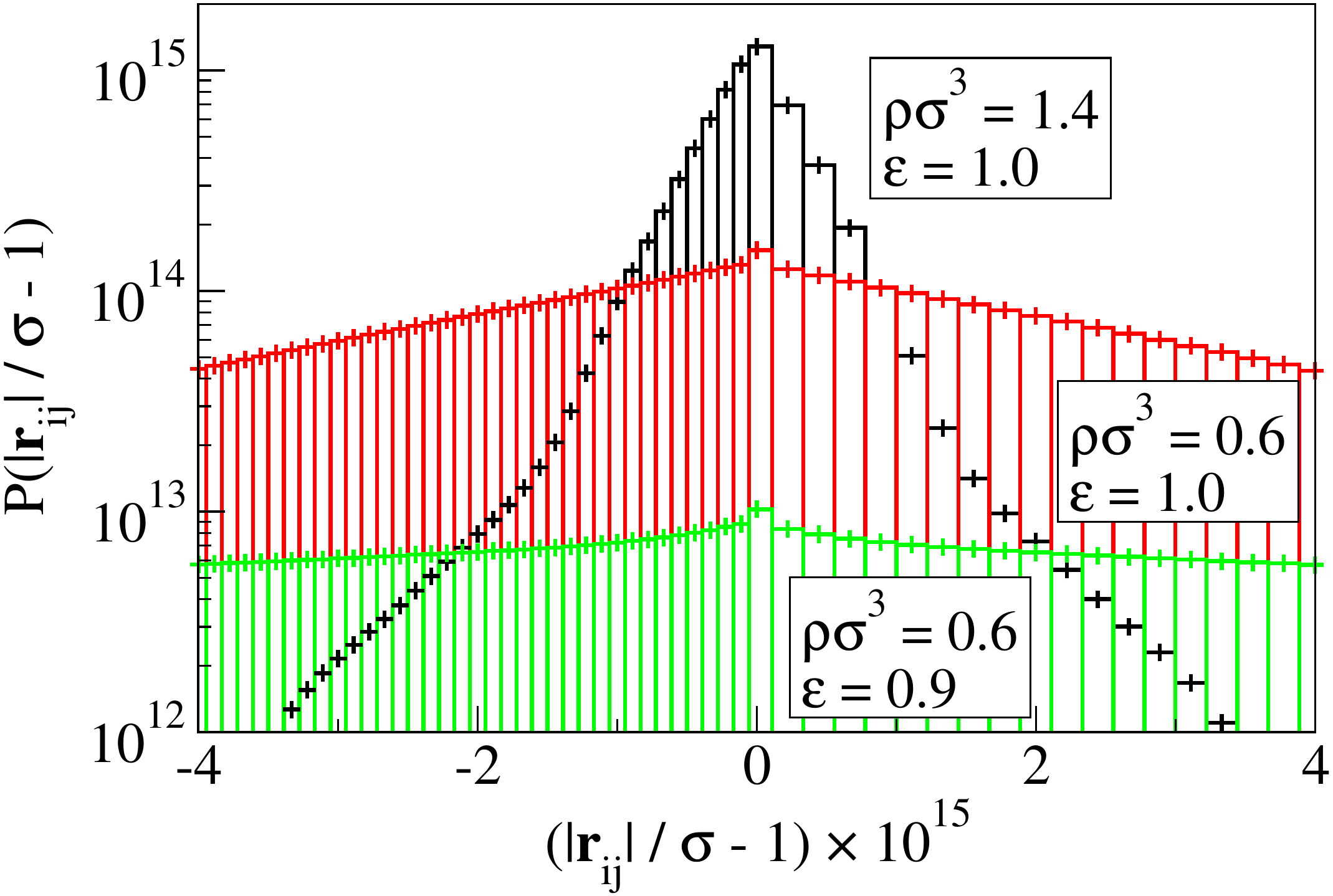}
  \end{center}
  \caption{\label{fig:overlaphist} Histograms of the impact
    separations for a simulation of $N=13\,500$ hard spheres under
    periodic boundary conditions, collected over $10^9$
    collisions. Each bar represents a single floating point number,
    marked at the top, and the range of continuous values that it
    represents due to the finite precision. Data for different
    elasticities, $\varepsilon$, and densities $\rho=N/V$, where $V$
    is the primary image volume are presented as seperate
    histograms. There is a precision change at zero due to a change in
    the floating-point exponent when crossing
    $\left|\vec{r}_{ij}\right|=\sigma$. }
\end{figure}

The overlapping case is problematic as the particles have numerically
entered the infinite energy hard-core. To illustrate the magnitude of
these errors, simulations of $N=13\,500$ hard spheres under periodic
boundary conditions were performed and histograms of the measured
impact distances, collected over $10^9$ collisions, are given in
Fig.~\ref{fig:overlaphist}. The mean and mode separation on impact
correspond to the interaction diameter, $\sigma$, but it is clear that
particle separations on impact can fall on either side of this value.
The overlapping states are relatively minor and typically resolve
themselves in elastic systems as the particle pair is receding after
the impact; however, Fig.~\ref{fig:overlaphist} clearly demonstrates
that the infinite energy core of the potential is numerically
accessible and the magnitude of these errors are significantly
increased in inelastic systems. The overlapped states may degenerate
if one of the overlapping particles interacts before the overlap is
cleared, effectively causing a three-body impact which is in direct
conflict with the physical model of instantaneous collisions for hard
spheres.

While errors resulting in $\left|\vec{r}_{ij}\right|\gtrsim \sigma$
are uncritical for the stability of the algorithm, the opposite case,
$\left|\vec{r}_{ij}\right|\lesssim \sigma$ is fatal since the
computation of the next event corresponds to a collision where
$\vec{v}_{ij}\cdot\hat{\vec{r}}_{ij} > 0$. Execution of this collision
approximates entangled circular rings and this entanglement will
persist forever, unless it is resolved due to another numerical error
in a future collision.
%
As shown in Fig.~\ref{fig:overlaphist}, this situation is not a rare
event but concerns approximately half of all collisions, therefore,
{\em any} useful EDPD algorithm must provide measures to cope with
this situation.  There are two different approaches: a) avoid
situations where $\left|\vec{r}_{ij}\right|< \sigma$, and b) admit
such situations but provide methods to recover from them. One method
exploiting solution a) to prevent overlaps from forming due to
numerical errors is to retrospectively search along the pre-collision
trajectory of colliding particle pairs for a collision state which is
not overlapping~\cite{Allen_etal_1989}. Unfortunately, this iterative
approach is expensive and will limit the computational speed of EDPD.
A more efficient scheme to prevent overlaps by biasing the movement of
particles by (temporarily) modifying the particle diameter so that
detected interactions occur at a small distance from the invalid state
is proposed by P\"oschel and Schwager~\cite{Poschel_Schwager_2005}.
Unortunately, this approach does not exactly simulate the desired
system but a slightly different system following a slightly different
dynamics. It may also be shown that there are cases where these
approaches fail to prevent overlaps
forming~\cite{Poschel_Schwager_2005}.

\begin{figure}[htp]
  \begin{center}
    \includegraphics[width=0.5\columnwidth,clip]{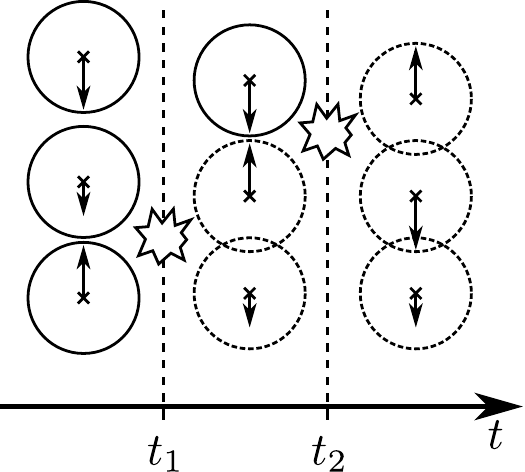}
  \end{center}
  \caption{An illustration of two impacts leading to three overlapping
	particles. This situation is difficult to resolve using simple
	collision detection algorithms. All overlaps in this sketch are
	strongly exaggerated for improved visibility.}
  \label{fig:three-spheres}
\end{figure}

\begin{figure}[htp]
  \begin{center}
    \includegraphics[width=0.5\columnwidth,clip]{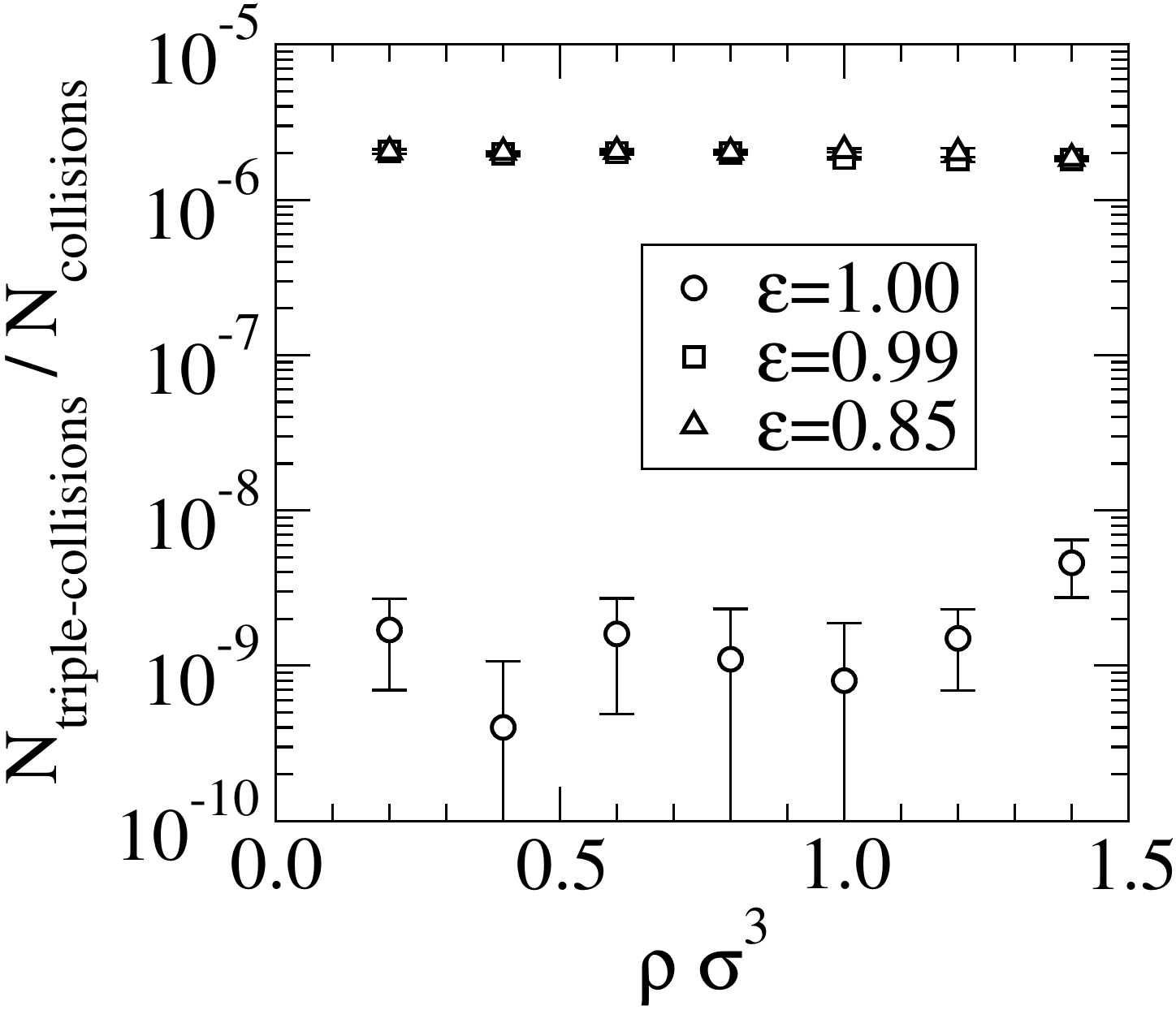}
  \end{center}
  \caption{The frequency of three-body collisions resulting in a doubly
	overlapped particle in hard-sphere simulations
	as a function of elasticity, $\varepsilon$, and reduced density.}
  \label{fig:three-body-impact}
\end{figure}

Specific discussions on how to handle event detection for overlapping
particle pairs corresponding to solution b) are
common~\cite{Allen_etal_1989,Bannerman_etal_2011b,Donev_etal_2005a,Frenkel_Maguire_1983,Guttenberg_2011,McNamara_etal_2000,Reichardt_Wiechert_2007};
however, these approaches often either implicitly disable interactions
between overlapped pairs, admit significant overlaps to form before
attempting to correct them, or rewind simulation time in an
uncontrolled manner, all of which fail to resolve three-body
contacts/events.
A one-dimensional illustration of a sequence of impacts which leads to
a three-body contact is given in
Fig.~\ref{fig:three-spheres}. Initially, two particles impact and
overlap due to numerical errors in the event calculation. This
situation typically resolves quickly as the particles are receding
from each other, but in rare cases a third particle may impact the
overlapping pair of particles before the overlap is resolved.  This
leads to a three-body contact where there are overlapping particles
which are approaching each other. Figure~\ref{fig:three-body-impact}
presents measurements on the frequency of these events in hard-sphere
systems with different densities and elasticities. In dilute elastic
systems ($\varepsilon=1$), these three-body events are extremely rare
($\approx10^{-9}$ three-body events per event processed) which may
explain why they are not discussed in the literature as published
simulations are rarely this long. In inelastic systems, clustering
effects~\cite{McNamara_Young_1994} and partial collapse events
increase the frequency of three-body events by three orders of
magnitude, making it likely that several will occur during a single
simulation run. If these three-body interactions cause overlapped
particle pairs to approach, simple algorithmic implementations for
hard-sphere event detection~\cite{Alder_Wainwright_1959,Haile_1997}
will return negative values of $\Delta t$ for overlapped
pairs. Execution of these events will cause the simulation to perform
an unchecked rewind, leading to other particle pairs to overlap,
particularly at high densities. It is clear that, even for simple
discrete-potentials like hard spheres, a stable simulation algorithm
must account for these numerical errors and their consequences,
particularly for inelastic systems and long simulation times.

\section{\label{sec:improvedalgorithm}Stable EDPD for hard-sphere systems}

\subsection{\label{sec:general-approach}General approach}

A general approach for devising stable event-detection algorithms
requires the introduction of an {\em overlap function} and the concept
of {\em stabilizing interactions}. Overlap functions are common in
studies on event-driven asymmetric-potential
systems~\cite{Allen_etal_1989}, where they have also been called the
overlap potential~\cite{Donev_etal_2005b} or the indicator
function~\cite{vanZon_Schofield_2008}.  For a simple hard-sphere
system with zero relative acceleration, a suitable overlap function,
$f_\text{HS}$, is defined via Eq.~\eqref{eq:hs-quadratic}:
\begin{equation}
  \label{eq:hs-overlapfunc}
  f_\text{HS}(t + \Delta t) = \Delta t^2\,\vec{v}_{ij}^{\,2} + 2\,\Delta
  t\,\vec{v}_{ij}\cdot\vec{r}_{ij} + \vec{r}_{ij}^{\,2} - \sigma^2
\end{equation}
where again all variables are evaluated at the current time $t$. The
overlap function is a function of time which characterizes the
relative position of particles moving along certain trajectories with
respect to overlap. Typically it is proportional to the distance or
squared distance between the closest points on the surfaces of the two
tested objects. The overlap function is negative for particle pairs in
an invalid state, and positive or zero in all valid states. In this
sense, it is a penalty function for invalid states.  By definition,
the overlap function transforms the search for event times into a
search for the roots of the overlap function, $f(t)$. In addition, the
sign of the overlap function can be used as a test if the particle
pair is in an invalid state. Crucially, these properties allow the
derivative of the overlap function to be used as a test if an invalid
state ($f(t)< 0$) is either improving or stable in time
($\dot{f}(t)\ge0$) or not ($\dot{f}(t)<0$). For the hard-sphere case,
the derivative is given by the following expression
\begin{equation}
  \label{eq:hs-overlapfunc-deriv}
  \dot{f}_\text{HS}(t + \Delta t) = 2\,\Delta t\,\vec{v}_{ij}^{\,2} +
  2\,\vec{v}_{ij}\cdot\vec{r}_{ij}\,.
\end{equation}

A stabilizing interaction is a collision that is performed by the
algorithm {\em immediately} after a preceding collision to ensure that
overlapping particles do not approach one another. It is generated in
response to a negative and decreasing overlap function and prevents
the overlap function decaying any further.

Using the overlap function and the concept of stabilizing
interactions, a stable algorithm for hard-sphere systems can be
defined as an algorithm which ensures the overlap function does not
decrease for any particle pair in contact or in an overlapping state:

{\em When testing for collisions between a pair of hard spheres at a
  time, $t$, consider the overlap function, $f_\text{HS}$. A collision
  occurs after the smallest non-negative time interval, $\Delta t$,
  that satisfies the following condition:}
\begin{align}
  \left(\vphantom{\dot{f}_\text{HS}} f_\text{HS}\left(t+\Delta t\right)\le 0\right) \text{{ }and }
  \left(\dot{f}_\text{HS}\left(t+\Delta t\right)<0\right)
\end{align}
For glancing interactions (see Fig.~\ref{fig:hard_sphere_roots}c),
corresponding to degenerate roots of Eq.~\eqref{eq:hs-overlapfunc}, no
sign change in $\dot{f}_\text{HS}$ occurs. The definition of the
stable algorithm will allow the particles to come in to contact but,
without a sign change in the derivative, no event will be scheduled
and no impulse will be applied. This is acceptable for models without
friction/rotation but it is a point of ambiguity in the implementation
of models with tangential forces. Here we recommend that glancing
interactions are also excluded from generating impulses in systems
with friction as there are no forces in the normal direction of the
contact.

\subsection{\label{sec:stable-HS} Algorithmic implementation for hard-sphere systems}

An implementation of a stable event detection algorithm for
hard-sphere collisions, in accordance with the definition in the
previous section, is presented in
Algorithm~\ref{alg:spherein}. Unsurprisingly, this algorithm is
similar to previously published
algorithms~\cite{Alder_Wainwright_1959,Haile_1997,Poschel_Schwager_2005}
for detecting hard-sphere collisions; however, it differs by the
addition of the second if-statement on
Line~\ref{alg:sphereinadditional}. Only the earliest root of the
overlap function generates interactions, as only the earliest root of
the overlap function has a decreasing overlap function (marked with a
filled circle in Fig.~\ref{fig:hard_sphere_roots}a). The calculation
of this root suffers from catastrophic
cancellation~\cite{Press_etal_1986} and so the alternate form of the
quadratic equation must be used, as suggested by P\"oschel and
Schwager~\cite{Poschel_Schwager_2005}.
Algorithm~\ref{alg:spherein} shows no predefined precision threshold;
hence it could be used directly with arbitrary~\cite{Granlund_2013} or
exact~\cite{Mehlhorn_etal_2013} precision libraries to reduce the
number of stabilizing collisions at the expense of more complicated
dynamic data structures and significantly longer computation times.

\begin{algorithm}
  \tcc{Ensure that the trajectory is before the minimum in $f$.}%
  \lIf{$\vec{v}_{ij}\cdot\vec{r}_{ij} \ge 0$}{ \Return $\infty$\; }
  \tcc{Catch overlapped and approaching states.}%
  \lIf{$\vec{r}^2_{ij}-\sigma^2 \le 0$}{%
    \label{alg:sphereinadditional} \Return $t$\; } \tcc{Catch misses
    (Fig.~\ref{fig:hard_sphere_roots}b) and glancing impacts
    (Fig.~\ref{fig:hard_sphere_roots}c).}%
  \lIf{$\left[(\vec{v}_{ij}\cdot\vec{r}_{ij})^2 -
      \vec{v}_{ij}^2(\vec{r}^2_{ij} - \sigma^2)\right]\le0$}{%
    \Return $\infty$\;%
  }%
  \tcc{Calculate first root of $f$ (see
    Fig.~\ref{fig:hard_sphere_roots}a).}%
  $\Delta t \longleftarrow \left(\vec{r}^2_{ij} -
    \sigma^2\right)/\left(-\vec{v}_{ij}\cdot\vec{r}_{ij} +
    \sqrt{(\vec{v}_{ij}\cdot\vec{r}_{ij})^2 -
      \vec{v}_{ij}^2(\vec{r}^2_{ij}-\sigma^2)}\right)$\;%
  \Return $t+\Delta t$\;
  \caption{The stable EDPD algorithm for collision detection between
	two hard spheres, $i$ and $j$, with collision diameter $\sigma$ as
	depicted in Fig.~\ref{fig:hard_sphere_roots}.}
  \label{alg:spherein}
\end{algorithm}

\begin{figure}[htp]
  \begin{center}
    \includegraphics[width=0.75\columnwidth,clip]{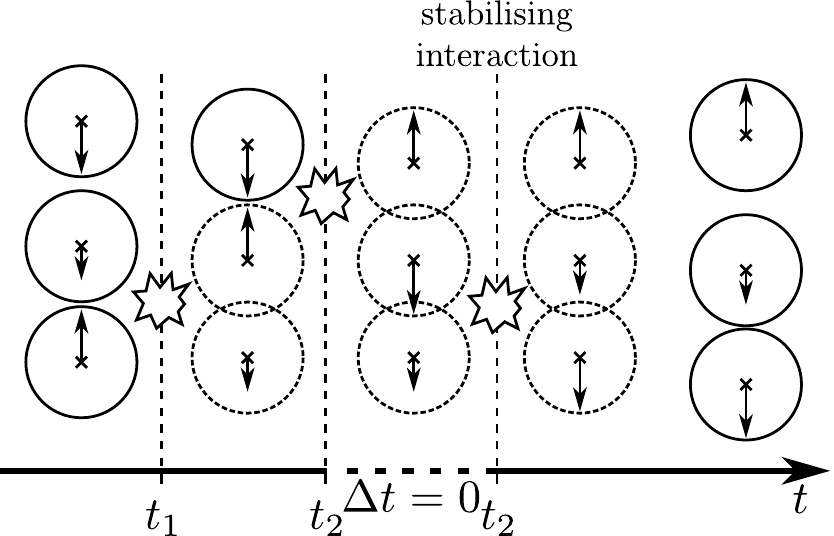}
  \end{center}
  \caption{An illustration of two impacts leading to three overlapping
    particles. This situation is difficult to resolve; however, if a
    third ``stabilizing'' collision is executed between the two
    initially-overlapping particles, all particles will begin to move
    apart and the overlaps will clear after a short time. Note that
	although the directions of the velocities of the lower two particles
	are not changed by the stabilizing collision, the direction of the
	relative velocity is.}
  \label{fig:three-spheres-resolved}
\end{figure}

Fig.~\ref{fig:three-spheres-resolved} sketches how this stable
algorithm would resolve the interactions between the particles for the
example system introduced in Sec.~\ref{sec:eventerrors}. The critical
step is the stabilizing collision occurring between the lower two
particles immediately after the execution of the first event at time
$t_2$. It is this event which ensures the system moves towards a valid
state, as shown in the last segment of
Fig.~\ref{fig:three-spheres-resolved}. At no point are the overlaps,
introduced by numerical error, allowed to increase in time during the
free motion of the system. Thus, the stable algorithm proposed here
handles three-body events by ensuring the overlap function never
deteriorates.

Although the stable algorithm defines the dynamics for three-body
interactions, it should not be used as a physical model for these
effects. It only provides a stable definition of the dynamics for rare
cases where multiple events occur at the same time (due to numerical
errors). In systems where three-body collisions or persistent contacts
are common, an appropriate model must be used. This model may also be
event-driven (e.g., stepped potentials~\cite{Thomson_etal_2014}) but
this again requires the general approach outlined here to ensure that
the simulation is stable with respect to numerical errors.

There are some cases where the execution of collisions will not
cause the overlap function to increase. Fortunately, this behavior is
often required to recover the correct dynamics, as illustrated in the
following section.

\section{\label{sec:bouncingball} Advanced example: Bouncing ball}

\begin{figure}
  \begin{center}
    \includegraphics[width=0.7\columnwidth,clip]{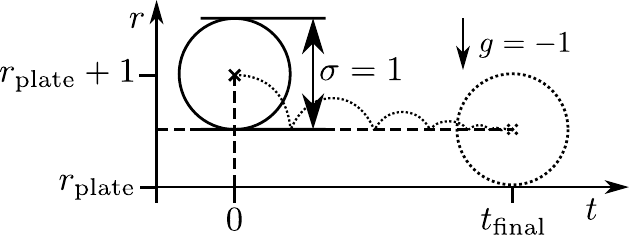}
  \end{center}
  \caption{\label{fig:bouncingball} An illustration of the position $r$
    of a ball of diameter $\sigma=1$ as it falls from a height
    $r(t=0)=r_{\mathrm{plate}}+1$ onto a plate located at
    $r_{\mathrm{plate}}$, coming to rest after a time
	$t_{\mathrm{final}}$.}
\end{figure}

A simple system to extend and test the applicability of the stable
algorithm introduced for hard spheres is the one-dimensional system of
an inelastic hard sphere falling under the influence of gravity onto a
hard plate, as described in Fig.~\ref{fig:bouncingball}.  The exact
solution to this system is that the sphere comes to rest after an
infinite number of impacts in a finite time~\cite{Falcon_etal_1998},
known as an inelastic collapse~\cite{McNamara_Young_1994}.

Attempting to numerically simulate the inelastic collapse highlights
the dramatic effect that small precision errors can have. In the
example explored here, the sphere is initialized with zero velocity,
$v=0$, and positioned above the plate. The time until the next impact,
$\Delta t$, is calculated from the largest positive root of the
following overlap function:
\begin{align}
  \label{eq:ballplateoverlap}
  f_{\mathrm{ball}\to\mathrm{plate}}(t+\Delta t) &= r+\Delta
  t\,v+\frac{\Delta t^2}{2}\,g - r_{\mathrm{plate}} -
  \frac{\sigma}{2}
\end{align}
where $r$ and $v$ are the position and velocity of the particle of
diameter $\sigma$ at time $t$, $r_{plate}$ is the position of the hard
plate, and $g$ is the gravitational acceleration. Approaching this
naively, the event time is given by the quadratic formula
\begin{align}
  \label{eq:bounceballdeltat}
  \Delta t &= -\frac{v+\sqrt{v^2 - 2\, g\,[r - r_{\mathrm{plate}} -
  \sigma/2]}}{g}\,.
\end{align}
Although the current position of the sphere relative to the plane is
different for the calculation of the first and all later impact times,
there is no qualitative difference between these two cases. For the
calculation of the first impact time the sphere is at a height of
$r_{\mathrm{plate}}+1$, while for all later impacts the current height
above the plate is $\sigma/2 + \delta$, where $\delta$ is a small
deviation due to numerical precision.  Once the next impact or
``event'' time is determined, the sphere is moved forward to the time
of the impact, its velocity is inelastically reflected,
$v'=-\varepsilon\,v$.  This process is then repeated until a
sufficient number of impacts has occurred or an error is
encountered. To explore the algorithm's stability, the origin of the
plate, $r_{\mathrm{plate}}$, is uniformly sampled using $10^6$ points
from the range $[0,1]$. All other parameters of this test are
presented in Fig.~\ref{fig:bouncingball} and a coefficient of
restitution of $\varepsilon=0.5$ is used.

The simple coordinate transformation of varying $r_{\mathrm{plate}}$
should yield identical results; however, the finite precision of the
floating point math causes some difficulties: For approximately $50\%$
of the sampled values of $r_{\mathrm{plate}}$, the simulation must be
halted after only $\approx30$ events on average as the argument of the
square root in Eq.~\eqref{eq:bounceballdeltat} becomes negative.  This
result arises due to the sphere overlapping the wall after a collision
by a small amount $\approx10^{-16}\,\sigma$, exactly as in the
hard-sphere system. The argument of the square root then turns
negative once the velocity decays to the point where the peak of the
trajectory no longer escapes the small overlap with the wall. A naive
implementation may assume that if there are no real roots to
Eq.~\eqref{eq:ballplateoverlap} the ball misses the plate; however,
this is clearly impossible in this system. This system demonstrates
that for inelastic systems with external forces, it is easy to enter
regions of undefined dynamics.

One possible treatment of this case is to halt the simulation
immediately; however, this will leave the system with a non-zero
velocity, a finite number of events in the trajectory, and an overlap.
Applying the stable algorithm defined in
Sec.~\ref{sec:general-approach} to the overlap function in
Eq.~\eqref{eq:ballplateoverlap} provides a more satisfactory
resolution where the motion of the particle is continued until it
reaches its peak, where the velocity reduces to zero.  The algorithm
also causes an infinite number of future events to take place at zero
time to prevent the overlap from increasing again, which approximates
the exact solution of inelastic collapse to the precision of the
calculations.  To complete the description of the stable algorithm in
this case, an optimized stable event-detection rule for a ball falling
onto a plane in three-dimensions is given in
Algorithm~\ref{alg:spherewall}.

\begin{algorithm}
  \lIf{$\hat{\vec{n}}\cdot\left(\vec{r}_i-\vec{r}_{wall}\right)<0$}{ $\hat{\vec{n}} \longleftarrow -\hat{\vec{n}}$\; }%
  $g\longleftarrow\hat{\vec{n}}\cdot\vec{g}$\;
  $r\longleftarrow\hat{\vec{n}}\cdot\left(\vec{r}_i-\vec{r}_{wall}\right)$\;
  $v\longleftarrow\hat{\vec{n}}\cdot\vec{v}_i$\;
  \lIf{$r\le\sigma/2$ {\bf and} $v<0$}{\Return $t$\;}%
  \If{$g=0$}{%
    \lIf{$v \ge 0$}{\Return $\infty$\;}%
    \Return $t-\left(r-\sigma/2\right)/v$\;
  }%
  $\Delta t_{min}\longleftarrow - v / g$\;%
  \eIf{$g<0$}{%
    \lIf{$v^2-2\,g(r-\sigma/2)<0$}{ \Return $t+\Delta t_{min}$\;}%
    $\Delta t_{1},\,\Delta t_{2}\longleftarrow\text{QuadraticFormula}\left(g\Delta t^2/2+v\Delta t+r - \sigma/2=0\right)$\;
    \Return $t+\max(\Delta t_{1},\,\Delta t_{2})$\;
  }{%
    \lIf{$(v^2-2\,g(r-\sigma/2)<0)$ {\bf or} $(\Delta t_{min}<0)$}{ \Return $\infty$\;}%
    $\Delta t_{1},\,\Delta t_{2}\longleftarrow\text{QuadraticFormula}\left(g\Delta t^2/2+v\Delta t+r - \sigma/2=0\right)$\;
    \Return $t+\min(\Delta t_{1},\,\Delta t_{2})$\;
  }
  \caption{The stable EDPD algorithm for collision detection between a
    particle $i$, under acceleration $\vec{g}$, with a plane defined by
    a normal $\hat{\vec{n}}$ and point $\vec{r}_{wall}$. The
    QuadraticFormula function returns the two real roots of the passed
    quadratic using the stable formulas~\cite{Press_etal_1986}.}
  \label{alg:spherewall}
\end{algorithm}

\section{\label{sec:conclusions}Conclusions}
A general and stable approach to event-detection in hard-sphere
systems has been proposed and sample implementations given in
Algorithms~\ref{alg:spherein} and \ref{alg:spherewall}. It defines the
dynamics of overlapping particles to minimize their effect on the
trajectory of the system. The algorithm relies on treating
interactions as stabilizing events for overlapping particle pairs
which are not improving with time.

Although this work has focussed on the hard-sphere system, the general
concept can be extended to the full range of particle systems studied
using event-driven techniques. The extension to other assymmetric
hard-core potentials is straightforward provided an overlap function,
$f$, with the characteristics outlined in
Sec.~\ref{sec:general-approach} and algorithms to determine its
derivative and roots are available. These are available in certain
cases~\cite{vanZon_Schofield_2008} but are non-trivial and must be
implemented numerically for some systems such as ellipsoidal
particles~\cite{Donev_etal_2005b}.

Algorithm~\ref{alg:spherewall} demonstrates that the stable algorithm
can be applied to boundary interactions. An extension of the algorithm
to sphere-triangle interactions would allow the simulation of complex
geometries in biological processes~\cite{Byrne_Waxham_Kubota_2010} and
the rapid design and import of boundaries from CAD programs for the
study of granular/solids processing
systems~\cite{Poschel_Schwager_2005}. The primary difficulty in this
extension is the definition of a suitable overlap function for
collision detection of composite objects.

To apply the technique to molecular systems, an extension to softer
stepped~\cite{Thomson_etal_2014} or terraced potentials is
required. Such potentials, including the fundamental square-well
potential, require additional care in the definition of the overlap
function as the invalid states of the model depend on the interaction
history of the particle pairs. The generalization to stepped
potentials, composite objects, and complex shapes will be explored in
future publications.

Reference implementations of all algorithms presented in this paper
are available in the open-source DynamO~\cite{Bannerman_etal_2011}
simulation package.

\begin{acknowledgements}
We gratefully acknowledge the support of the Cluster of Excellence
'Engineering of Advanced Materials' at the University of
Erlangen-Nuremberg, which is funded by the German Research Foundation
(DFG) within the framework of its 'Excellence Initiative'. The authors
would like to thank Roland Reichardt and J.\ Sebasti\'an Gonz\'ales for
discussing their EDPD implementations.
\end{acknowledgements}



\begin{thebibliography}{10}
\providecommand{\url}[1]{{#1}}
\providecommand{\urlprefix}{URL }
\expandafter\ifx\csname urlstyle\endcsname\relax
  \providecommand{\doi}[1]{DOI~\discretionary{}{}{}#1}\else
  \providecommand{\doi}{DOI~\discretionary{}{}{}\begingroup
  \urlstyle{rm}\Url}\fi

\bibitem{Alder_Wainwright_1959}
Alder, B.J., Wainwright, T.E.: Studies in molecular dynamics. 1. {G}eneral
  method.
\newblock J. Chem. Phys. \textbf{31}(2), 459--466 (1959).
\newblock \doi{10.1063/1.1730376}

\bibitem{Allen_etal_1989}
Allen, M.P., Frenkel, D., Talbot, J.: Molecular dynamics simulation using hard
  particles.
\newblock Comp. Phys. Reports \textbf{9}, 301--353 (1989).
\newblock \doi{10.1063/1.1730376}

\bibitem{Bannerman_etal_2011b}
Bannerman, M.N., Kollmer, J.E., Sack, A., Heckel, M., M\"uller, P., P\"oschel,
  T.: Movers and shakers: Granular damping in microgravity.
\newblock Phys. Rev. E \textbf{84}, 011,301 (2011).
\newblock \doi{10.1103/PhysRevE.84.011301}

\bibitem{Bannerman_etal_2011}
Bannerman, M.N., Sargant, R., Lue, L.: Dynamo: A free {O(N)} general
  event-driven simulator.
\newblock J. Comp. Chem. \textbf{32}, 3329--3338 (2011).
\newblock \doi{10.1002/jcc.21915}

\bibitem{Byrne_Waxham_Kubota_2010}
Byrne, M.J., Waxham, M.N., Kubota, Y.: Cellular dynamic simulator: An event
  driven molecular simulation environment for cellular physiology.
\newblock Neroinform.  (2010).
\newblock \doi{10.1007/s12021-010-9066-x}

\bibitem{Chapela_etal_2010}
Chapela, G.A., del Rio, F., Benavides, A.L., Alejandre, J.: Discrete
  perturbation theory applied to {L}ennard-{J}ones and {Y}ukawa potentials.
\newblock J. Chem. Phys. \textbf{133}, 234,107 (2010).
\newblock \doi{10.1063/1.3518711}

\bibitem{Cui_Elliott_2002}
Cui, J., Elliott, J.R.: Phase diagrams for a multistep potential model of
  $n$-alkanes by discontinuous molecular dynamics and thermodynamic
  perturbation theory.
\newblock J. Chem. Phys. \textbf{116}, 8625--8631 (2002).
\newblock \doi{10.1063/1.1469608}

\bibitem{DeltourBarrat:1997}
Deltour, P., Barrat, J.L.: Quantitative study of a freely cooling granular
  medium.
\newblock J. Physique I \textbf{7}, 137--151 (1997)

\bibitem{Donev_etal_2005a}
Donev, A., Torquato, S., Stillinger, F.H.: Neighbor list collision-driven
  molecular dynamics simulation for nonspherical hard particles. {I}.
  {A}lgorithmic details.
\newblock J. Comp. Phys. \textbf{202}, 737--764 (2005).
\newblock \doi{10.1016/j.jcp.2004.08.014}

\bibitem{Donev_etal_2005b}
Donev, A., Torquato, S., Stillinger, F.H.: Neighbor list collision-driven
  molecular dynamics simulation for nonspherical hard particles: {II}.
  {A}pplications to ellipses and ellipsoids.
\newblock J. Comp. Phys. \textbf{202}, 765--793 (2005).
\newblock \doi{10.1016/j.jcp.2004.08.025}

\bibitem{Dorfman_Ernst_1989}
Dorfman, J.R., Ernst, M.H.: Hard-sphere binary-collision operators.
\newblock J. Stat. Phys. \textbf{57}, 581--593 (1989).
\newblock \doi{10.1007/BF01022823}

\bibitem{Falcon_etal_1998}
Falcon, E., Laroche, C., Fauve, S., Coste, C.: Behavior of one inelastic ball
  bouncing repeatedly off the ground.
\newblock Eur. Phys. J. B \textbf{3}, 45--57 (1998).
\newblock \doi{10.1007/s100510050283}

\bibitem{Frenkel_Maguire_1983}
Frenkel, D., Maguire, J.F.: Molecular dynamics study of the dynamical
  properties of an assembly of infinitely thin hard rods.
\newblock Mol. Phys. \textbf{49}(3), 503--541 (1983).
\newblock \doi{10.1080/00268978300101331}

\bibitem{Granlund_2013}
Granlund, T., {the GMP development team}: {GNU MP}: {T}he {GNU} {M}ultiple
  {P}recision {A}rithmetic {L}ibrary, 5.1.3 edn. (2013).
\newblock \url{http://gmplib.org/}

\bibitem{Guttenberg_2011}
Guttenberg, N.: Approximate hard-sphere method for densely packed granular
  flows.
\newblock Phys. Re \textbf{83}, 051,306 (2011).
\newblock \doi{10.1103/PhysRevE.83.051306}

\bibitem{Haile_1997}
Haile, J.M.: Molecular Dynamics Simulation - Elementary Methods.
\newblock Wiley-Interscience, New York (1997)

\bibitem{Jefferson_1985}
Jefferson, D.R.: Virtual time.
\newblock TOPLAS \textbf{7}(3), 404--425 (1985).
\newblock \doi{10.1145/3916.3988}

\bibitem{Lubachevsky:1991}
Lubachevsky, B.D.: How to simulate billiards and similar systems.
\newblock J. Comp. Phys. \textbf{94}, 255--283 (1991)

\bibitem{Luding:1998}
Luding, S., McNamara, S.: How to handle the inelastic collapse of a dissipative
  hard-sphere gas with the tc model.
\newblock Granular Matter \textbf{1}, 113--128 (1998)

\bibitem{Marin:1993}
Marin, M., Risso, D., Cordero, P.: Efficient algorithms for many-body hard
  particle molecular-dynamics.
\newblock J. Comp. Phys. \textbf{109}, 306--317 (1993)

\bibitem{McNamara_etal_2000}
McNamara, S., Flekk\o{}y, E.G., M\aa{}l\o{}y, K.J.: Grains and gas flow:
  Molecular dynamics with hydrodynamic interactions.
\newblock Phys. Rev. E \textbf{61}, 4054 (2000)

\bibitem{McNamara_Young_1994}
McNamara, S., Young, W.R.: Inelastic collapse in two dimensions.
\newblock Phys. Rev. E \textbf{50}, R28--R31 (1994).
\newblock \doi{10.1103/PhysRevE.50.R28}

\bibitem{Mehlhorn_etal_2013}
Mehlhorn, K., N\"aher, S., Seel, M., Uhrig, C.: LEDA: A Platform for
  Combinatorial and Geometric Computing.
\newblock Cambridge University Press (1999)

\bibitem{Montaine:2011}
Montaine, M., Heckel, M., Kruelle, C., Schwager, T., P\"oschel, T.: Coefficient
  of restitution as a fluctuating quantity.
\newblock Phys. Rev. E \textbf{84}, 041,306 (2011)

\bibitem{Mueller:2011}
M\"uller, P., P\"oschel, T.: Collision of viscoelastic spheres: {C}ompact
  expressions for the coefficient of normal restitution.
\newblock Phys. Rev. E \textbf{84}, 021,302 (2011)

\bibitem{Mueller:2012}
M\"uller, P., P\"oschel, T.: Oblique impact of frictionless spheres: {O}n the
  limitations of hard sphere models for granular dynamics.
\newblock Gran. Matter \textbf{14}, 115--120 (2012)

\bibitem{Mueller:2013}
M\"uller, P., P\"oschel, T.: Event-driven molecular dynamics of soft particles.
\newblock Phys. Rev. E \textbf{87}, 033,301 (2013)

\bibitem{Paul_2007}
Paul, G.: A complexity ${O}(1)$ priority queue for event driven molecular
  dynamics simulations.
\newblock J. Comput. Phys. \textbf{221}(2), 615--625 (2007).
\newblock \doi{10.1016/j.jcp.2006.06.042}

\bibitem{Poschel_Schwager_2005}
P\"oschel, T., Schwager, T.: Computational Granular Dynamics.
\newblock Springer-Verlag, Berlin Heidelberg (2005).
\newblock \doi{10.1007/3-540-27720-X}

\bibitem{Press_etal_1986}
Press, W.H., Flannery, B.P., Teukolsky, S.A., Vetterling, W.T.: Numerical
  Recipes: The Art of Scientific Computing.
\newblock Cambridge University Press (1986)

\bibitem{Rapaport_1980}
Rapaport, D.C.: Event scheduling problem in molecular dynamics simulation.
\newblock J. Comput. Phys. \textbf{34}(2), 184--201 (1980).
\newblock \doi{10.1016/0021-9991(80)90104-7}

\bibitem{Reichardt_Wiechert_2007}
Reichardt, R., Wiechert, W.: Event driven algorithms applied to a high energy
  ball mill simulation.
\newblock Gran. Mat. \textbf{9}, 251--266 (2007).
\newblock \doi{10.1007/s10035-006-0034-y}

\bibitem{Schwager:2008}
Schwager, T., P\"oschel, T.: Coefficient of restitution for viscoelastic
  spheres: {T}he effect of delayed recovery.
\newblock Phys. Rev. E \textbf{78}, 051,304 (2008)

\bibitem{ShidaAnzai:1992}
Shida, K., Anzai, Y.: Reduction of the event-list for molecular dynamic
  simulation.
\newblock Comput. Phys. Comm. \textbf{69}, 317--329 (1992)

\bibitem{ShidaYamada:1995}
Shida, K., Yamada, S.: Reduced event-list on an array for many-body simulation.
\newblock Comput. Phys. Comm. \textbf{86}, 289--296 (1995)

\bibitem{Thomson_etal_2014}
Thomson, C., Lue, L., Bannerman, M.N.: Mapping continuous potentials to
  discrete forms.
\newblock J. Chem. Phys. \textbf{140}, 034,105 (2014).
\newblock \doi{10.1063/1.4861669}

\bibitem{Vahid_etal_2008}
Vahid, A., Sans, A.D., Elliot, J.R.: Correlation of mixture vapor-liquid
  equilibria with the {SPEADMD} model.
\newblock Ind. Eng. Chem. Res. \textbf{47}, 7955--7964 (2008).
\newblock \doi{10.1021/ie800374h}

\bibitem{vanZon_Schofield_2008}
van Zon, R., Schofield, J.: Event-driven dynamics of rigid bodies interacting
  via discretized potentials.
\newblock J. Chem. Phys. \textbf{128}, 154,119 (2008).
\newblock \doi{10.1063/1.2901173}

\end{thebibliography}
\end{document}